\documentclass[a4,12pt,epsf]{article}

\begin{document}
\begin{center}{\Large \bf
	$S_3$ Higgs Potential and Texture-zeros in Supersymmetric Standard Model 
	\footnote{talked by T. Shingai at the Summer Institute 2006, International Workshop on Supersymmetry, Electroweak Symmetry Breaking and Particle Cosmology, APCTP, Pohang, Korea, 23-30 August 2006.}}
\end{center}

\begin{center}
	Satoru Kaneko$^{1,}$\footnote{satoru@ific.uv.es}, Hideyuki Sawanaka$^{2,}$\footnote{hide@muse.sc.niigata-u.ac.jp}, Takaya Shingai$^{2,}$\footnote{shingai@muse.sc.niigata-u.ac.jp},

Morimitsu Tanimoto$^{3,}$\footnote{tanimoto@muse.sc.niigata-u.ac.jp} and Koichi Yoshioka$^{4,}$\footnote{yoshioka@higgs.phys.kyushu-u.ac.jp}
\vspace{6pt}\\

{\it
$^1$Physics Department, Ochanomizu University, Tokyo 112-8610, Japan}

{\it
$^2$Graduate school of Science and Technology, 
Niigata University, Niigata 950-2181, Japan
}

{\it
$^3$Physics Department, Niigata University, Niigata 950-2181, Japan}

{\it
$^4$Physics Department, Kyushu University, Fukuoka 812-8581, Japan}

\end{center}
\begin{abstract}
The mass matrix forms of quarks and leptons are discussed in theory
with permutation flavor symmetry. The structure of scalar potential is
analyzed in case that electroweak doublet Higgs fields have 
non-trivial flavor symmetry charges. We find that realistic forms of
mass matrices are obtained dynamically in the vacuum of the theory,
where some of Higgs bosons have vanishing expectation values which
lead to vanishing elements in quark and lepton mass matrices. 
An interesting point is that, due to
the flavor group structure, the up and down quark mass matrices are
automatically made different in the vacuum, which lead to
non-vanishing generation mixing. It is also discussed that flavor
symmetry is needed to be broken in order not to have too light
scalars.
\end{abstract}

\section{Introduction}
Texture-zeros, vanishing elements of fermion mass matrices, 
in quark and lepton mass matrix are successful to
predict masses and mixing angles. However the origin of zero is not 
clear, which is a motivation of our model. As an approach for this problem 
we consider the discrete flavor symmetry approach. This approach is studied by 
many authors{\cite{discrete symmetry}}. Flavor symmetry is expected to 
be a clue to understand the masses and mixing angles of quarks and leptons because
it reduces the number of free parameters in Yukawa couplings and some testable 
predictions of masses and mixing angles generally follow(see references in {\cite{our model}}).

A interesting point of our model is that dynamical realization of Texture-zeros in
the discrete flavor symmetry approach. In previous model, in order to derive 
Texture-zeros certain Yukawa couplings are forbidden by the discrete symmetry. 
In our model however we consider that some of Higgs vacuum expectation values(VEVs) vanish 
by electroweak symmetry breaking(EWSB) in flavor basis, that is, we consider multi-Higgs system and 
Texture-zeros are derived by EWSB dynamically. 

In our model we take $S_3$ symmetry, permutations of three objects, as the discrete symmetry. 
The reasons why we adopt $S_3$ are that this symmetry is the smallest group of non 
commutative discrete groups and $S_3$ has three 
irreducible representations, doublet $\bf 2$, singlet ${\bf 1_{S}}$, pseudo singlet 
${\bf 1_{A}}$, so that it is easy to assign the flavor symmetry representations to three generations such as
${\bf 2} + {\bf 1_{S}}$. In addition, we consider all the $S_{3}$ irreducible representations in this model.

\section{$S_3$ invariant mass matrix on supersymmetry}

In this section, the $S_3$ invariant mass matrices are presented{\cite{mass matrix}}. 
We consider supersymmetric theory and we suppose that two of three generations belong to $S_3$ doublets and the others are singlets. Using the following tensor product of $S_3$ doublet, $\phi^{c} = \sigma_{1} \phi^{*}= \sigma_{1}(\phi^{*}_{1},\phi^{*}_{2})^{T},\ \psi = (\psi_{1},\psi_{2})^{T}$,
\begin{equation} 
	\begin{array}{ccccccc}
		\phi^{c} \times \psi &=& (\phi_2 \psi_2,\phi_1 \psi_1)^{T} & + & (\phi_1 \psi_2 - \phi_2 \psi_1) & + & (\phi_1 \psi_2 + \phi_2 \psi_1), \\
		    & & {\bf 2} & & {\bf 1_{A}} & & {\bf 1_{S}} 
	\end{array}
\end{equation}
the $S_{3}$ invariant mass matrices are obtained as
\begin{equation}
   M_{D}=
      \left(
          \begin{tabular}{cc|c}
	      $aH_{1}$ & $b H_{S} + c H_{A}$ & $d H_{2}$ \\
	      $b H_{S} - c H_{A}$ & $a H_{2}$ & $d H_{1}$ \\
	      \hline
	      $e H_{2}$ & $e H_{1}$ & $ f H_{S}$
	        \end{tabular}
      \right),\qquad
		M_{R}=
        \left(
	    \begin{tabular}{cc|c}
		    & $M_{1}$ &   \\
	      $M_{1}$ &   &  \\
	      \hline
	             &    &  $M_{2}$
	    \end{tabular}
	\right),
\end{equation}
where $a, b, \cdots,f$ are independent Yukawa coupling constants, $M_1, M_{2}$ are majorana masses. Now we assign ${\bf 1_{S}}$ to third generation temporarily. We reconfigure this assignment later. 

\section{$S_3$ invariant Higgs scalar potential analysis}
In our model we consider the following eight Higgs bosons,
\begin{equation}
	H_{uS},H_{dS},H_{uA},H_{dA},H_{u1},H_{d1},H_{u2},H_{d2}.
\end{equation}
Our purpose in this section is to discuss whether or not there are some vacuum patterns with no parameter relations in terms of vanishing-VEVs. 
Therefore we have to analyze eight equations at vacuum, which correspond to each Higgs field.
\begin{table}[t]
\begin{minipage}{0.5 \textwidth}
	\begin{center}
\begin{tabular}{|c|c|c|c|c|c|c|c|} \hline
$v_{uS}$ & $v_{dS}$ & $v_{uA}$ & $v_{dA}$ & $v_{u1}$ & $v_{u2}$ 
& $v_{d1}$ & $v_{d2}$ \\ \hline \hline
 $0$ & $0$ & $0$ & $0$ & $0$ &     &     & $0$ \\ \hline
 $0$ & $0$ & $0$ & $0$ &     & $0$ & $0$ &     \\ \hline
 $0$ & $0$ & $0$ & $0$ &     &     &     &     \\ \hline
 $0$ & $0$ &     &     & $0$ & $0$ & $0$ & $0$ \\ \hline
 $0$ & $0$ &     &     & $0$ &     &     & $0$ \\ \hline
 $0$ & $0$ &     &     &     & $0$ & $0$ &     \\ \hline
 $0$ & $0$ &     &     &     &     &     &     \\ \hline
 \end{tabular}
 \end{center}
 \end{minipage}
 \begin{minipage}{0.5 \textwidth}
\begin{center}
\begin{tabular}{|c|c|c|c|c|c|c|c|} \hline
$v_{uS}$ & $v_{dS}$ & $v_{uA}$ & $v_{dA}$ & $v_{u1}$ & $v_{u2}$ 
& $v_{d1}$ & $v_{d2}$ \\ \hline \hline
     &     & $0$ & $0$ & $0$ & $0$ & $0$ & $0$ \\ \hline
     &     & $0$ & $0$ & $0$ &     &     & $0$ \\ \hline
     &     & $0$ & $0$ &     & $0$ & $0$ &     \\ \hline
     &     & $0$ & $0$ &     &     &     &     \\ \hline
     &     &     &     & $0$ & $0$ & $0$ & $0$ \\ \hline
     &     &     &     & $0$ &     &     & $0$ \\ \hline
     &     &     &     &     & $0$ & $0$ &     \\ \hline
\end{tabular}
\end{center}
\end{minipage}
\caption{All possible minima of the scalar potential for $S_3$ singlet
and doublet Higgs fields without tuning of Lagrangian parameters for
electroweak symmetry breaking. The blank entries denote
non-vanishing VEVs.}
{\label{summary}}
\end{table} 
In general, these equations are the coupled equations through a common parameter which contains all the Higgs VEVs. However we can separate these equations into three parts for the singlet
, the pseudo singlet
and the doublet
because vanishing-VEVs makes the equations trivial within each sector. 
Therefore analyzing the equations
, possible 14 VEV patterns(Table {\ref{summary}}) are obtained with no parameter relations in terms of vanishing-VEVs. 

\section{Quark and lepton mass textures}

In previous section, we got 14 VEV patterns.
Now let us analyze these patterns phenomenologically. At first we can 
obtain the most interesting pattern of 14 patterns. 
This pattern is the following.
\begin{equation}
	v_{uS}=v_{dS}=v_{u1}=v_{d2}=0,\qquad v_{uA},v_{dA},v_{u2},v_{d1}\not= 0. {\label{vev}}
\end{equation}
This pattern leads to the simplest texture(i.e.the maximal number of zero matrix elements) with non-trivial flavor mixing. Next we consider mass matrices obtained from this VEV pattern. We only took ${\bf 2} + {\bf 1_S}$ as the $S_3$ representations of three generation matter fields so that the $S_3$ charge assignments of matter fields has a complexity. For example we can assign ${\bf 1_{S}}$ to any generation in general. As results of exhausting all the $S_3$ charge assignments for the quark sector and assuming $SU(5)$ grand unification{\cite{su(5)GUT}}{\cite{georgi jarlskog}} for the lepton sector, mass matrices and predictions of this model are derived as 
 
\begin{equation}
	M_u =
		\left(
			\begin{array}{ccc}
					& b_u &		\\
				d_u &		& f_u \\
				 & - f_u & i_u
			\end{array}
		\right),\qquad
	M_d =
		\left(
			\begin{array}{ccc}
					& b_d &		\\
				d_d &	e_d	&  \\
				-e_d &  & i_d
			\end{array}
		\right), \qquad
	M_e =
		\left(
			\begin{array}{ccc}
					& d_d &	3e_d	\\
				b_d &	-3 e_d	& \\
				 &  & i_d
			\end{array}
		\right),
\end{equation}
\begin{equation}
	M_{\nu} =
		\left(
			\begin{array}{ccc}
				& b_{\nu} &	c_{\nu}	\\
				-b_{\nu} &	e_{\nu}	&  \\
				g_{\nu} & & 
			\end{array}
		\right),\qquad
	M_R =
		\left(
			\begin{array}{ccc}
					& M_1 &		\\
				M_1 &		&  \\
				 &  & M_2
			\end{array}
		\right),
\end{equation}
\begin{equation}
	\left| V_{cb} \right| = \sqrt{\frac{m_c}{m_t}},\qquad \left| V_{e3} \right| \ge 0.04{\cite{PDG}}.
\end{equation}
where blank entries denote zero and each parameter in $M_{u},M_{d},M_{e},M_{\nu}$ such as $d_{u},d_{d},b_{\nu}$ denote a product of a Yukawa coupling and a VEV, for example $d_{u} = d v_{u2}$.
\section{Higgs mass spectrum and $S_3$ soft breaking in B-term}

The $S_3$ potential has an enhanced global symmetry $SU(2) \times U(1)^2$ and leads to massless Nambu-Goldstone bosons 
in the electroweak broken phases. It is therefore reasonable to softly break the flavor symmetry within the scalar potential.
We introduce the following supersymmetry-breaking soft terms which do not break phenomenological characters of the exact $S_3$ model.
\begin{equation}
	V_{\not S_3} = b_{SD}H_{uS}H_{d2}+ b'_{SD}H_{u1}H_{dS}+ b_{AD}H_{uA}H_{d1}+ b'_{AD}H_{u2}H_{dA} + {\rm h.c.}
\end{equation}
These soft terms have not only the same phenomenological characters as exact $S_3$ model but also a character which we can take the same VEV pattern as ({\ref{vev}}) in previous section with no parameter relations.

\section{Tree level FCNC}

Since there are multiple electroweak doublet Higgs bosons which couple to 
matter fields, flavor-changing processes are mediated at classical 
level by these Higgs fields.	We can show that all but one have masses of the order of supersymmetry breaking parameters. Therefore the experimental observations of FCNC rare events would lead to a bound on the supersymmetry breaking scale.
Among various experimental constraints, we find the most important constraint comes from the neutral K meson mixing. For the heavy mass eigenstates, the tree-level $K_{L}-K_{S}$ mass difference $\Delta m^{\rm tree}_K$ is given by the matrix element of the effective Hamiltonian between K mesons{\cite{FCNC}}.
$\Delta m^{\rm tree}_K$ contains $M_{H}$, which is an average of the Higgs masses $1/M^{2}_{H} = \frac{1}{4}\left( 1/M^{2}_{H^{0}_{1}} + 1/ M^{2}_{H^{0}_{2}} + 1/M^{2}_{H^{0}_{3}}+1/M^{2}_{H^{0}_{4}} \right)$, and a free parameter $\eta$, which contains the down type quark Yukawa couplings.
It is found that heavy Higgs masses are bounded from below so as to suppress the extra Higgs contribution compared with the standard model one, which bound is roughly given by
\begin{equation}
  M_{H} \ge
	    \left\{
			    \begin{array}{cl}
						3.8 {\rm TeV} &(\eta = 0) \\
						1.4 {\rm TeV} &(\eta = 0.03) \\
					\end{array}
				\right.
,
\end{equation}
where we take $\eta = 0,\ 0.03$ as typical values.

\section{Summary}

In our model we have discussed the structure of Higgs potential and fermion mass 
matrices in supersymmetric models with $S_3$ flavor symmetry and examined possible 
vanishing elements of quark and lepton mass matrices. As results of exhausting the 
patterns of flavor symmetry charges of matter fields, some predictions such that 
the lepton mixing $V_{e3}$ is within the range which will be tested in near future 
experiments are obtained. We have also discussed the physical mass spectrum of 
Higgs bosons and the tree level FCNC processes which is propagated by heavy Higgs fields. 
From the tree level FCNC process analysis, it is found that heavy Higgs masses, 
which is the order of soft supersymmetry breaking scale, are a few TeV.

\vspace*{12pt}
\noindent
{\bf Acknowledgement}
\vspace*{6pt}

\noindent
The Summer Institute 2006 is sponsored by the Asia 
Pacific Center for Theoretical Physics and the BK 21 
program of the Department of Physics, KAIST.
We would like to thank the organizers of Summer Institute 2006 and thank J. Kubo, T. Kobayashi and H. Nakano for helpful discussions.



\begin{thebibliography}{99}
	\bibitem{discrete symmetry}
		J. Kubo, A. Mondragon, M. Mondragon and E. Rodriguez-Jauregui, Prog. Theor. Phys. {\bf 109} (2003) 795 [Erratum-ibid. {\bf 114} (2005) 287]; \\
		T. Kobayashi, J. Kubo and H. Terao, Phys. Lett. B {\bf 568} (2003) 83;\\
		J. Kubo, H. Okada and F. Sakamaki, Phys. Rev. D {\bf 70} 036007.
	\bibitem{our model}
		S. Kaneko, H. Sawanaka, T. Shingai, M. Tanimoto and K. Yoshioka, arXiv:hep-ph/0609220.
	\bibitem{mass matrix}
		N. Haba and K. Yoshioka, Nucl. Phys. B {\bf 739} (2006) 254, 
	  [aiXiv:hep-ph/0511108].	
	\bibitem{su(5)GUT}
		H. Georgi and S.L. Glashow, Phys. Rev. Lett. {\bf 32} (1974) 438.
	\bibitem{georgi jarlskog}
		H. Georgi and C. Jarlskog, Phys. Lett. B {\bf 88} (1979) 297; \\
		S. Dimopoulos, L.J. Hall and S. Raby, Phys. Rev. Lett. {\bf 68} (1992) 1984.
	\bibitem{PDG}
		S. Eidelman {\it et} {\it al}. [Particle Data Group Collaboration], Phys. Lett. B {\bf 592} (2004) 1.
		\bibitem{FCNC}
		F. Gabbiani, E. Gabrielli, A. Masiero and L. Silvestrini,
		Nucl. Phys. B {\bf 477} (1996) 321.
\end{thebibliography}
\end{document}